# Nonequilibrium Solvent Polarization Effects in Real-Time Electronic Dynamics of Solute Molecules Subject to Time-Dependent Electric Fields: A New Feature of the Polarizable Continuum Model

Gabriel Gil,[\*,†,‡] Silvio Pipolo,[¶] Alain Delgado,[§] Carlo Andrea Rozzi,[∥] and Stefano Corni[\*,†,∥]

[†]Dipartimento di Scienze Chimiche, Università degli studi di Padova, via F. Marzolo 1, 35131 Padova, Italia

[‡]Instituto de Cibernética, Matemática y Física, Calle E esq 15 Vedado 10400, La Habana, Cuba

[¶]Université de Lille, CNRS, Centrale Lille, ENSCL, Université d'Artois UMR 8181—UCCS Unité de Catalyse et Chimie du Solide, F-59000, Lille, France

[§]Xanadu, 777 Bay Street, Toronto, Ontario M5G 2C8, Canada

[∥]Istituto Nanoscienze—CNR, via Campi 213/A, 41125 Modena, Italia

**ABSTRACT:** We develop an extension of the time-dependent equation-of-motion formulation of the polarizable continuum model (EOM-TDPCM) to introduce nonequilibrium cavity field effects in quantum mechanical calculations of solvated molecules subject to time-dependent electric fields. This method has been implemented in Octopus, a state-of-the-art code for real-space, real-time time-dependent density functional theory (RT-TDDFT) calculations. To show the potential of our methodology, we perform EOM-TDPCM/RT-TDDFT calculations of *trans*-azobenzene in water and in other model solvents with shorter relaxation times. Our results for the optical absorption spectrum of *trans*-azobenzene show (i) that cavity field effects have a clear impact in the overall spectral shape and (ii) that an accurate description of the solute shape (as the one provided within PCM) is key to correctly account for cavity field effects.

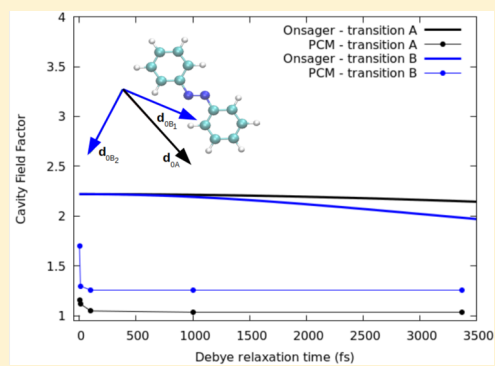

## 1. INTRODUCTION

The electronic and optical properties of a molecule embedded in a dielectric environment, such as a solvent, can be substantially different from those of its gas-phase counterpart. The presence of the molecule induces a polarization in the surrounding that in turn interacts with the molecule itself via the so-called reaction field. Additionally, if the molecule is subject to an external electric field, the effective field acting locally on the molecule (namely, the cavity field) contains another polarization contribution of the environment due to the applied field.[1] Hence, when simulating a solvated molecule interacting with any electromagnetic perturbation (e.g., light), both the reaction and cavity field effects must be considered.

A suitable framework to account for polarization effects in quantum-mechanical calculations of molecules surrounded by a dielectric environment is the polarizable continuum model (PCM).[2−5] Within PCM, (i) the environment is regarded as a continuum and infinite dielectric medium featuring a frequency-dependent dielectric function; (ii) the molecule is treated explicitly, that is, by considering its atomic structure, and generally within quantum mechanics; (iii) a void cavity hosts the molecule, conforming with its shape and separating it from the dielectric environment by a sharp interface; and (iv) the reaction- and cavity-field polarization of the medium are described in terms of apparent surface charge (ASC) densities sitting on the cavity surface that are induced by the molecule and the applied electric fields, respectively.

The PCM formulation can accommodate nonequilibrium effects in the molecule-environment dynamics when shifting from the electrostatic[2,5] to the time-dependent quasi-electrostatic[3,6−9] regime. However, until recently, nonequilibrium features were only accessible through the sudden excitation approximation, where the molecular density changes abruptly from the ground to the excited state. In response to this sudden excitation, the solvent degrees of freedom are crudely divided in two groups: the "slow" ones that are fully unable to adapt to the molecular density rearrangements, and the "fast" ones that follow instantaneously the molecular density evolution. Although this slow/fast splitting of the environment polarization contributions represents a qualitatively correct scheme to treat nonequilibrium polarization effects, it neglects the fact that even the slow environment polarization (albeit dominated by its inertial character) has a finite relaxation time and can as well adjust to the molecular excitation in the long term.[2]

Currently, a computationally convenient time-dependent extension of PCM (TDPCM) has made possible to study the real-time evolution of the molecular density subject to time-dependent quasi-electrostatic perturbations, together with the polarization of the external medium.[10−13] Such a TDPCM scheme introduces polarization fields that are history-dependent and not instantaneously in equilibrium with the molecule or the applied fields. These nonequilibrium polarization effects arise ultimately from the frequency-dependent dielectric function describing the environment, and they are physically linked to the fact that the environment takes a finite amount of time to react in response to the electrostatic perturbations, that is, the molecular density fluctuations and the applied fields. Mathematically, this formulation can be cast in terms of a set of equations of motion (EOM) for the ASCs dictating the evolution of the medium polarization in real-time. In the case of solvated molecules, EOM-TDPCM has been formulated for the reaction field;[13] however, a similar account for the cavity field is still missing.

On a related note, even though treating the reaction field within the quasi-electrostatic regime is generally pertinent, an equivalent description of the cavity field does not apply, in principle, when a molecule surrounded by a dielectric medium is irradiated with light. To overcome this limitation, in ref 14, cavity field effects have been rigorously tackled within classical electrodynamics, by using the full set of Maxwell equations to describe the cavity modification of electromagnetic fields traveling in a medium when impinging a vacuum cavity. Notwithstanding, under convenient gauges and the relevant long-wavelength limit,[14,28] the electrodynamics result has been brought to match the standard PCM account of the cavity field effects,[3−5] where an effective dipole of the embedded molecule interacts with the time-dependent electric field of the light wave.

The aim of this paper is to present an extension of EOM-TDPCM to consider nonequilibrium cavity field polarization effects for molecules embedded in an homogeneous dielectric medium, such as a solvent. This contribution represents a step further in considering more realistic solvent polarization effects. Nonequilibrium cavity field effects may impact, strongly and nontrivially, the electronic and optical properties obtained from real-time quantum mechanical simulations of solvated molecules, especially when at least some of the solvent degrees of freedom relax in a short time scale (0.1−1 of ps). Cavity field effects treated within the PCM framework enable us to consider in a more realistic manner complex molecules, where the solute shape departs strongly from symmetric shapes (such as a sphere or an ellipsoid) and furnishes quantitative an qualitative improvement over some standard descriptions for the cavity field, such as Onsager's model.

To show the potential of our methodology, we perform real-space real-time time-dependent density functional theory (TDDFT) simulations of solvated *trans*-azobenzene to assess the influence of the cavity field in the optical spectrum and time-dependent properties. The molecule of choice does not have spherical symmetry and is largely anisotropic. As a consequence, some of its relevant excited states feature transition dipoles oriented along different directions. *trans*-Azobenzene is, therefore, a good candidate to emphasize the nontrivial role of cavity field effects in changing the optical spectrum shape. We present results for the molecule immersed in water. Furthermore, to highlight the role of nonequilibrium cavity field effects, we provide comparison between the results obtained for the molecule solvated in water and in different fictitious solvents with faster orientational relaxation.

The present account is organized as follows. In section 2, we recall the PCM theory of cavity field effects for a molecule embedded in a dielectric environment. Section 2.1 is devoted to the electrostatics formulation of cavity field effects. In section 2.2, we extend the latter to tackle the case of time-dependent electric fields in the long-wavelength limit and using the length gauge for the case of a molecule in solution. Section 2.3 introduces time-dependent quantum mechanics for electrons in a solvated molecule subject to an external electric field. Section 2.4 focuses on a specific time-dependent quantum-mechanics treatment, namely, real-time TDDFT (RT-TDDFT). In section 2.5, we obtain the expression for the full photoabsorption cross-section spectrum for solvated molecules, emphasizing cavity-field contributions and the ways to account for them through real-time propagation of an effective molecular dipole. Sections 2.2 and 2.5 are specific for the case of solvent, while the rest are rather general for any medium considered. In section 3, we present the computational details (section 3.1) and the results for the optical absorption spectra (section 3.2) of *trans*-azobenzene in water. Section 3.2 focuses on the cavity field and nonequilibrium cavity field effects in the optical absorption spectrum of *trans*-azobenzene. Finally, in section 4, we summarize the relevant points of the Article, and we stress the potential of our methodology for future and more accurate studies of molecules in solution.

## 2. PCM THEORY OF CAVITY FIELD EFFECTS

**2.1. Electrostatics.** Let us consider a static electric field applied to the solution. Far from the vacuum cavity hosting the solute, the dielectric representing the solvent is homogeneous and the electric field there is the *Maxwell field*, $\mathbf{F_M}$.[15] Close to and within the solute cavity, the electric field is locally modified by the presence of the cavity itself. Any distribution of charges placed inside the cavity (e.g., a molecule) is affected not only by the Maxwell field as if it were applied in an homogeneous medium but also by the medium polarization because of the cavity boundary. The effects of such electric field contribution supplementing the Maxwell field are called cavity field effects. To treat them within a PCM framework, one defines an effective electrostatic potential.[14]

Such effective applied potential (that includes cavity field effects) acting on any distribution of charges, $v_C(\mathbf{r})$ to keep the nomenclature of ref 14, is the solution of a generalized Laplace problem, where the dielectric constant is position-dependent, with boundary conditions at the cavity surface and far from it (see refs 1, 2, and 5). The latter generalized Laplace problem can be equivalently written as a generalized Poisson problem, considering a fictitious distribution of charges $\rho_M(\mathbf{r})$ placed outside the vacuum cavity (however far from it) as the source of the Maxwell field. The PCM framework,[2] in particular, in its integral equation formalism (IEF) approach[16] allow us to solve this kind of Poisson problem. Ref 16 shows an analogous IEF-PCM formulation in the reaction field case that we followed thoroughly to obtain an analogous account for the cavity field case.

As it is customary within PCM framework, we split the solution of the aforementioned Poisson problem into two contributions, namely,[17]

$$v_C(\mathbf{r}) = v_M(\mathbf{r}) + v_{cf}(\mathbf{r}) \qquad (1)$$

However, the specific partition of the potential we employ differs from the usual choice in the case of the reaction field.

Here, the first contribution, $v_M(\mathbf{r})$, is the potential associated with the Maxwell field, whereas the remaining term, $v_{cf}(\mathbf{r})$, represents a polarization potential that solves the Laplace equation inside and outside the cavity and fulfills the boundary conditions from the original Poisson problem. Thence, $v_{cf}(\mathbf{r})$ is a so-called layer potential that can be written in terms of ASC density $\sigma_{cf}(\mathbf{r})$ located at the interface $\Gamma$, that is

$$v_{cf}(\mathbf{r}) = \int_\Gamma \frac{\sigma_{cf}(\mathbf{s})}{|\mathbf{r} - \mathbf{s}|} dA(\mathbf{s}) \tag{2}$$

where $dA(\mathbf{s})$ is the surface area element corresponding to $\mathbf{s} \in \Gamma$.[2,16]

Within the numerical strategy known as boundary element method (BEM), we consider a discrete (but fine) tessellation of the cavity surface and approximate the latter as[2]

$$v_{cf}(\mathbf{r}) \approx \sum_{j=1}^{N_{tess}} \frac{\sigma_{cf}(\mathbf{s}_j)A_j}{|\mathbf{r} - \mathbf{s}_j|} = \sum_{j=1}^{N_{tess}} \frac{q_{cf}^j}{|\mathbf{r} - \mathbf{s}_j|} \tag{3}$$

where $N_{tess}$ is the number of elements (tesserae) in $\Gamma$'s tessellation, whereas $\mathbf{s}_j$, $A_j$, and $q_{cf}^j$ are the position, the surface area and the polarization point charge associated with the $i$th tessera.

IEF-PCM equations for the polarization charges $\{q_{cf}^j\}$ in terms of the potential that induces them can be written as[2,16]

$$q_{cf}^j = \sum_{k=1}^{N_{tess}} Q_{cf}(\mathbf{s}_j, \mathbf{s}_k) v_M(\mathbf{s}_k) \tag{4}$$

or in matrix form, $\mathbf{q}_{cf} = \mathbb{Q}_{cf} \mathbf{v}_M$. $\mathbb{Q}_{cf} = \{Q_{cf}(\mathbf{s}_j, \mathbf{s}_k)\}$, with $j$, $k$ = 1, ..., $N_{tess}$, is the PCM response matrix depending on the cavity shape, the tessellation of its surface, and the dielectric constant of the medium.[2,16]

To get the cavity field PCM response matrix starting from our partition 1, we follow an analogous derivation to that laid down in ref 16 to obtain the reaction-field PCM response matrix. It yields

$$\mathbb{Q}_{cf} = -[\mathbb{S}_e(2\pi\mathbb{I} + \mathbb{D}_i^*\mathbb{A}) + (2\pi\mathbb{I} - \mathbb{D}_e\mathbb{A})\mathbb{S}_i]^{-1}$$
$$\times [\mathbb{S}_e\mathbb{S}_i^{-1}(2\pi\mathbb{I} + \mathbb{D}_i\mathbb{A}) - (2\pi\mathbb{I} + \mathbb{D}_e\mathbb{A})] \tag{5}$$

where

$$\mathbb{A} = \{A_j \delta_{jk}\}$$

$$\mathbb{S}_i = \left\{ \frac{1}{|\mathbf{s}_j - \mathbf{s}_k|} \right\}$$

$$\mathbb{D}_i = \left\{ \frac{(\mathbf{s}_j - \mathbf{s}_k) \cdot \mathbf{n}(\mathbf{s}_k)}{|\mathbf{s}_j - \mathbf{s}_k|^3} \right\} \tag{6}$$

with $j$, $k$ = 1, ..., $N_{tess}$, and $\mathbf{n}(\mathbf{s}_k)$ is a unitary normal vector to the $k$-th tessera, pointing outward from the cavity. Matrices $\mathbb{S}_{i/e}$ and $\mathbb{D}_{i/e}$ are the BEM representation of Calderon operators in the IEF.[16] Heretofore, we have introduced very general definitions and, in particular, we have not assumed that the dielectric constant for the external medium is position-independent, as it is for a typical solvent. Indeed any in-homogeneous external dielectric medium, with a position-dependent dielectric constant of its own, can be tackled in the same way provided that the $\mathbb{S}_e$ and $\mathbb{D}_e$ matrices are known, which is to say that the Green's function of the external generalized Poisson problem were found.[2,16] Particularizing to the solvent case, the PCM response matrix takes the form

$$\mathbb{Q}_{cf} = \mathbb{S}_i^{-1} \left( 2\pi \frac{(\epsilon + 1)}{(\epsilon - 1)}\mathbb{I} - \mathbb{D}_i\mathbb{A} \right)^{-1} (2\pi\mathbb{I} + \mathbb{D}_i\mathbb{A}) \tag{7}$$

The cavity field PCM response matrices 5 and 7 differ from their reaction-field counterparts by two sign changes: one to the full expression and another one affecting the term $\mathbb{D}_i\mathbb{A}$ inside the second (noninverted) bracket; both "−" for reaction-field and "+" for the cavity-field case (cf., eq 3 in ref 13). PCM response matrices for the reaction- and cavity-field cases are different because the source charges inducing the medium polarization are located in different regions of the space in each case, that is, inside the vacuum cavity for the reaction field and outside (in the external medium) for the cavity field.

Partition 1 is not the only starting point to solve the original generalized Laplace problem for the applied electrostatic potential within the PCM framework. In Appendix A.1, we develop, for a different partition, an alternative derivation to cavity field effects within PCM.

## 2.2. Electrodynamics in the Long-Wavelength Limit.

In the previous section, we considered the static polarization response of a dielectric medium with a vacuum cavity due to an externally applied electrostatic field. Here, we study the dynamic polarization response of the medium because of a time-dependent electromagnetic field. This can be, for example, the electromagnetic field associated with a monochromatic light pulse probing the molecule in a specific spectroscopy.

Far from the cavity, in the solvent bulk, the electric field is again that proper for an homogeneous medium and is defined as the Maxwell field. Close to the molecule, we work in the long-wavelength limit and under the length-gauge,[14] so that locally the scalar Maxwell potential can be written as

$$v_M(\mathbf{r}, \omega) = -\mathbf{r} \cdot \mathbf{F}_M(\omega) \tag{8}$$

in the frequency domain. Within these assumptions, we may solve the generalized Poisson problem in the frequency domain, considering the complex dielectric function dependence on the frequency, that is, $\epsilon(\mathbf{r}; \omega)$. We solve the latter by means of the PCM eq 4 and the polarization potential (eq 3).

Let us focus on an external medium with the characteristics of a solvent (i.e., position-independent dielectric constant). In such a case, the PCM response matrix $\mathbb{Q}_{cf}(\omega)$ is known in terms of the frequency-dependent dielectric function of the external medium (eq 7 with $\epsilon \to \epsilon(\omega)$). Hence, the PCM eq 4 holds in the frequency domain as $\mathbf{q}_{cf}(\omega) = \mathbb{Q}_{cf}(\omega)\mathbf{v}_M(\omega)$.[2,13] To obtain the time-dependence of the polarization charges, we have to Fourier transform the latter PCM equation, that is,[13,18]

$$\mathbf{q}_{cf}(t) = \int_{-\infty}^{\infty} \mathbb{Q}_{cf}(t - t')\mathbf{v}_M(t')dt'$$

$$\mathbb{Q}_{cf}(t - t') = \frac{1}{2\pi} \int_{-\infty}^{\infty} \mathbb{Q}_{cf}(\omega)e^{-i\omega(t-t')} d\omega \tag{9}$$

The former expression is central in this work and reflects the physical meaning of the nonequilibrium cavity field effects. It shows that cavity field polarization charges—and therefore, the cavity field—depend upon the full previous history of the Maxwell potential and not just on its values at the current time.

*Debye EOM for the Cavity−Field Polarization Charges.* In the case of a solvent described by Debye's dielectric model, the evolution of cavity field polarization charges—governed by the latter expressions—can be reformulated in terms of an EOM. Such EOM was derived by Corni, Pipolo, and Cammi in ref 13 for the reaction-field polarization charges. It is also valid for the cavity field polarization charges once the molecular potential is replaced by the Maxwell electric field and the proper propagation matrices are obtained from the new form of the PCM response matrix, see eq 7.

We recall that Debye's model represents the dielectric medium by a set of freely rotating point-like dipoles whose orientation is determined by random collisions and by the applied electric field.[19] Therefore, Debye's model captures the polarization response of the medium to an electric perturbation due to the orientational relaxation of the solvent molecules. Within Debye's model, the external medium is characterized by the complex dielectric function

$$\epsilon(\omega) = \epsilon_d + \frac{(\epsilon_0 - \epsilon_d)}{1 - i\omega\tau_D} \tag{10}$$

where $\tau_D$ is the so-called Debye relaxation time, which is a physical measure of how fast the dielectric environment can adapt to the changes of the potential it is subject to. Mathematically, it interpolates between the static and dynamic dielectric constants in the limit of small and large relaxation time, $\epsilon_0$ and $\epsilon_d$, respectively.

The resulting EOM for the cavity field polarization charges in the Debye case is

$$\dot{\mathbf{q}}_{cf}(t) = \mathbb{Q}_{cf,d}\dot{\mathbf{v}}_M(t) + \tilde{\mathbb{Q}}_{cf,0}\mathbf{v}_M(t) - \mathbb{R}\mathbf{q}_{cf}(t) \tag{11}$$

$$\begin{cases} \mathbb{Q}_{cf,0/d} = -\mathbb{S}^{-1/2}\mathbb{K}_{cf,x}\mathbb{T}^\dagger\mathbb{S}^{-1/2} \\[6pt] \mathbb{K}_{cf}(\epsilon) = -\left[\frac{(\epsilon+1)}{(\epsilon-1)}2\pi\mathbb{I} - \Lambda\right]^{-1}(2\pi\mathbb{I} + \Lambda) \\[6pt] \mathbb{K}_{cf,0/d} = \mathbb{K}_{cf}(\epsilon_{0/d}) \\[6pt] \tilde{\mathbb{Q}}_{cf,0/d} = -\mathbb{S}^{-1/2}\mathbb{T}\tau^{-1}\mathbb{K}_{cf,0/d}\mathbb{T}^\dagger\mathbb{S}^{-1/2} \\[6pt] \tau_{ii} = \tau_D\frac{(2\pi - \Lambda_{ii})\epsilon_d + 2\pi + \Lambda_{ii}}{(2\pi - \Lambda_{ii})\epsilon_0 + 2\pi + \Lambda_{ii}} \\[6pt] \mathbb{S}^{-1/2}\mathbb{D}\mathbb{A}\mathbb{S}^{1/2} = \mathbb{T}\Lambda\mathbb{T}^\dagger \\[6pt] \mathbb{R} = \mathbb{S}^{-1/2}\mathbb{T}\tau^{-1}\mathbb{T}^\dagger\mathbb{S}^{1/2} \end{cases} \tag{12}$$

where $\Lambda$ and $\tau$ are diagonal matrices. The only changes in the EOM propagation matrices for the cavity field with respect to reaction-field case (cf., ref 13) are located in the definition of $\mathbb{K}_{cf}(\epsilon)$ in eq 12; more explicitly, they are sign changes (i) to the full expression ("+" for "−"), and (ii) to the matrix $\Lambda$ in the noninverted bracket ("−" for "+"). For a propagation with a finite Maxwell field switched-on at the initial time $t = 0$, $\mathbf{q}(0) = \mathbf{0}$. If we consider a medium with an infinite relaxation time, we get

$$\mathbf{q}_{cf}(t) = \mathbb{Q}_{cf,d}\mathbf{v}_M(t) \tag{13}$$

The case of a Dirac delta time-dependence is special and it deserves a separated treatment (see Appendix A.4).

Finally, the time-dependent equivalent of eq 3 is

$$v_{cf}(\mathbf{r}, t) \approx \sum_{j=1}^{N_{tes}} \frac{q_{cf}^j(t)}{|\mathbf{r} - \mathbf{s}_j|} \tag{14}$$

## 2.3. Time-Dependent Quantum Mechanics.
The time-dependent effective electronic Hamiltonian of a molecule embedded in a dielectric environment and interacting with a probing electric field in the long-wavelength limit, using the length gauge and within the PCM framework is[14]

$$\hat{H}(t) = \hat{H}_{\overline{mol}}(t) + \hat{V}_C(t) \tag{15}$$

$\hat{H}_{\overline{mol}}(t) = \hat{H}_{mol} + \hat{V}_{rf}(t)$ is the Hamiltonian of the embedded molecule in absence of external fields, collecting the isolated molecule (in vacuo) Hamiltonian and the interaction with the polarization of the surrounding medium due to the molecule itself (reaction field). On the other hand, $\hat{V}_C(t) = \hat{V}_M(t) + \hat{V}_{cf}(t)$ describes the interaction with the total electrostatic field acting on the molecule, including the interaction with the bare external field in the dielectric medium (Maxwell field) and with the polarization of the medium due to the applied electrostatic field (cavity field term). As long as the nuclei are fixed, the isolated molecule Hamiltonian $\hat{H}_{mol}$ is time-independent. The time-dependent interaction Hamiltonian can be written as

$$\hat{V}_x(t) = \sum_{i=1}^{N_e} v_x(\mathbf{r}_i, t), \text{ with } x = rf, M, cf \tag{16}$$

where $N_e$ and $\{\mathbf{r}_i\}$ are the number of electrons of the molecule and their position operators, respectively. Notice that the time-dependent reaction-field potential, $v_{rf}(\mathbf{r},t)$, depends on the time-dependent molecular potential

$$v_{mol}(\mathbf{s}, t') = \int_{\mathbb{R}^3} \frac{\rho_{mol}(\mathbf{r}, t')}{|\mathbf{r} - \mathbf{s}|}d\mathbf{r} \tag{17}$$

at *any* earlier time $t' < t$.[2,13] These time-dependencies can be taken into account altogether by means of an EOM for polarization charges equivalent to that in eq 11 that only depends on the molecular potential at current times.[13] In eq 17, $\rho_{mol}(\mathbf{r},t)$ is the time-dependent molecular charge density, collecting electronic and nuclei contributions.[2]

## 2.4. RT-TDDFT.
In this Article, the quantum-mechanical description of the molecule embedded in a dielectric medium and interacting with an external electric field (including reaction and cavity field effects) is tackled at the level of DFT[20] and TDDFT.[21] Particularly, we use the Kohn−Sham (KS) approach within DFT,[22] justified by the Hohenberg−Kohn theorems,[23] and its time-dependent extension within TDDFT, justified by the Runge-Gross[24] and the van Leeuwen[25] theorems.

We start from the time-independent KS equations[20]

$$\hat{h}[\rho_0(\mathbf{r})]\psi_i(\mathbf{r}) = \varepsilon_i\psi_i(\mathbf{r}) \tag{18a}$$

$$\rho_0(\mathbf{r}) = -\sum_{i=1}^{N_e} |\psi_i(\mathbf{r})|^2 \tag{18b}$$

where

$$\hat{h}[\rho_0] = \hat{h}_{mol}[\rho_0] + v_{rf}[\rho_0](\mathbf{r}) + v_C(\mathbf{r}) \tag{19}$$

$\{\psi_i(\mathbf{r})\}$ and $\{\varepsilon_i\}$ are the KS Hamiltonian, molecular orbitals and energies, respectively, whereas $\rho_0(\mathbf{r})$ is the ground-state electronic charge density. $\hat{h}_{mol}[\rho_0]$ is the KS Hamiltonian for the isolated molecule (in vacuo and in the absence of external fields) including the kinetic energy operator, nuclear attraction, Hartree, exchange and correlation potentials.

The time-dependent KS equations,[21]

$$\hat{h}[\rho(\mathbf{r}, t)](t)\psi_i(\mathbf{r}, t) = i\frac{\partial\psi_i(\mathbf{r}, t)}{\partial t} \tag{20a}$$

$$\rho(\mathbf{r}, t) = -\sum_{i=1}^{N_e} |\psi_i(\mathbf{r}, t)|^2 \tag{20b}$$

where

$$\hat{h}[\rho](t) = \hat{h}_{mol}[\rho] + v_{rf}[\rho](\mathbf{r}, t) + v_C(\mathbf{r}, t) \tag{21}$$

and $\{\psi_i(\mathbf{r},t)\}$ are the time-dependent KS Hamiltonian and molecular orbitals, respectively, whereas $\rho(\mathbf{r},t)$ is the time-dependent electronic charge density. The adiabatic approximation for the exchange and correlation term of KS Hamiltonian is usually assumed, that is to say that $\hat{h}_{mol}(t)$ is a functional of the density at the current time, that is, $\hat{h}_{mol}[\rho(\mathbf{r},t)]$.[21] However, the reaction-field term, $v_{rf}[\rho](\mathbf{r},t)$, is instead intrinsically diabatic and it depends on the density at earlier times.[2]

Provided that the solution for the starting time $t = 0$ is known, the propagation scheme[26] solves eq 20a. Within RT-TDDFT schemes using the adiabatic approximation, the initial KS orbitals are obtained from the time-independent KS equation (see eq 18a), that is, $\psi_i(\mathbf{r}, 0) = \psi_i(\mathbf{r})$.

## 2.5. Optical Absorption Spectrum of a Solvated Molecule.
In this section, we find the photoabsorption cross-section of a solvated molecule, aiming at including and singling-out cavity field effects. In addition, we assume an electromagnetic plane wave impinging on the system as the source of photoabsorption processes. Furthermore, we assume to be in the long-wavelength limit (within the length-gauge) and in the linear regime.[27]

The intensity of a monochromatic electromagnetic radiation with frequency $\omega$ traveling in a nonmagnetic medium is[28]

$$I(\omega) = \frac{c}{8\pi}|\mathbf{F}_M(\omega)|^2 n(\omega) \tag{22}$$

where $\mathbf{F}_M(\omega)$ is the amplitude of the Maxwell field and $n(\omega)$ is the refraction index of the medium. We recall that $n(\omega) = \sqrt{(|\epsilon(\omega)| + Re[\epsilon(\omega)])/2}$. Within our framework of assumptions, the rate of energy dissipation of a monochromatic electromagnetic radiation applied to the system can be written as[28]

$$P(\omega) = \frac{1}{2}\omega Im\left[\int d\mathbf{r}\rho_{mol}^*(\mathbf{r}, \omega)v_C(\mathbf{r}, \omega)\right] \tag{23}$$

where $\rho_{mol}(\mathbf{r},\omega)$ is the charge density of the molecule induced by the total potential $v_C(\mathbf{r},\omega)$, both in the frequency domain (i.e., Fourier transformed from $\rho_{mol}(\mathbf{r},t)$ and $v_C(\mathbf{r},t)$, respectively), collecting the Maxwell and cavity field terms. Finally, the photoabsorption cross-section is the ratio between eqs 23 and 22, yielding

$$\bar{\Sigma}(\omega) = \frac{4\pi}{c}\frac{\omega Im\left[\int d\mathbf{r}\rho_{mol}^*(\mathbf{r}, \omega)v_C(\mathbf{r}, \omega)\right]}{|\mathbf{F}_M(\omega)|^2 n(\omega)} \tag{24}$$

Using partition 1 and $v_M(\mathbf{r},\omega) = -\mathbf{F}_M(\omega)\cdot\mathbf{r}$ in eq 24, we get

$$\bar{\Sigma}(\omega) = \Sigma_M(\omega) + \Sigma_{cf}(\omega)$$

$$= -\frac{4\pi\omega}{c}\frac{Im[\mathbf{d}_{mol}^{cf*}(\omega)\cdot\mathbf{F}_M(\omega)]}{|\mathbf{F}_M(\omega)|^2 n(\omega)}$$

$$+ \frac{4\pi\omega}{c}\frac{Im\left[\int_{\mathbb{R}^3}\rho_{mol}^*(\mathbf{r}, \omega)v_{cf}(\mathbf{r}, \omega)d\mathbf{r}\right]}{|\mathbf{F}_M(\omega)|^2 n(\omega)} \tag{25}$$

where $\mathbf{d}_{mol}^{cf}(\omega)$ is the dipole of the molecule induced by the total potential in the frequency domain. By incorporating a superscript "cf", we emphasize that the latter includes implicitly cavity field effects, but it is not the only source of them in the cross-section spectrum. We now focus on the term $\Sigma_{cf}(\omega)$, which is the contribution to the cross-section coming from the polarization of the medium because of the Maxwell field and, henceforth, accounts explicitly for further cavity field effects. Using the expression of the potential $v_{cf}(\mathbf{r},\omega)$ in terms of its sources $\sigma_{cf}(\mathbf{r},\omega)$, we can write

$$\Sigma_{cf}(\omega) = \frac{4\pi\omega}{c}\frac{Im\left[\int_\Gamma v_{mol}^*(\mathbf{s}, \omega)\sigma_{cf}(\mathbf{s}, \omega)dA(\mathbf{s})\right]}{|\mathbf{F}_M(\omega)|^2 n(\omega)} \tag{26}$$

where $v_{mol}(\mathbf{s},\omega)$ is the Fourier transform of the quantity defined in eq 17. The polarization charge density due to the Maxwell field, $\sigma_{cf}(\mathbf{s},\omega)$, is a linear function of the Maxwell field, $\mathbf{F}_M$; therefore, by applying Euler theorem for linear functions, we get[14]

$$\Sigma_{cf}(\omega) = -\frac{4\pi\omega}{c}\frac{Im[\tilde{\mathbf{d}}^*(\omega)\cdot\mathbf{F}_M(\omega)]}{|\mathbf{F}_M(\omega)|^2 n(\omega)} \tag{27}$$

where

$$\tilde{\mathbf{d}}(\omega) = -\int_\Gamma v_{mol}(\mathbf{s}, \omega)\frac{\partial\sigma_{cf}(\mathbf{s}, \omega)}{\partial\mathbf{F}_M(\omega)}dA(\mathbf{s}) \tag{28}$$

Using eq 27, we may write the complete cross-section as

$$\bar{\Sigma}(\omega) = -\frac{4\pi\omega}{c}\frac{Im[\bar{\mathbf{d}}_{mol}^*(\omega)\cdot\mathbf{F}_M(\omega)]}{|\mathbf{F}_M(\omega)|^2 n(\omega)} \tag{29}$$

where $\bar{\mathbf{d}}_{mol}(\omega) = \mathbf{d}_{mol}^{cf}(\omega) + \tilde{\mathbf{d}}(\omega)$ is the effective induced dipole of the solvated molecule (external dipole in Onsager's nomenclature[1]). We recall that the cross-section of a molecule without cavity field effects can be obtained from eq 24 replacing $v_C(\mathbf{r},\omega)$ by $v_M(\mathbf{r},\omega)$, yielding

$$\Sigma(\omega) = -\frac{4\pi\omega}{c}\frac{Im[\mathbf{d}_{mol}^*(\omega)\cdot\mathbf{F}_M(\omega)]}{|\mathbf{F}_M(\omega)|^2 n(\omega)} \tag{30}$$

where $\mathbf{d}_{mol}(\omega)$ is the molecular dipole induced by the Maxwell field.

Notice that we can write the induced dipoles as

$$\mathbf{d}_{mol}(\omega) = \boldsymbol{\alpha}^{dd}(\omega)\mathbf{F}_M(\omega) \tag{31a}$$

$$\bar{\mathbf{d}}_{mol}(\omega) = \boldsymbol{\alpha}^{\bar{d}d}(\omega)\mathbf{F}_M(\omega) \tag{31b}$$

where $\boldsymbol{\alpha}(\omega)$ is the frequency-dependent first-order polarizability tensor. The first superscript labels the observable under examination and the second superscript keeps track of the magnitude coupled with the perturbator.[21,29] In eqs 31a and 31b, we monitor different observables, namely, the induced molecular and effective dipole (second superscript $d$ and $\bar{d}$, respectively), considering different interaction Hamil-tonians as a perturbation, that is, $\hat{V}_M(t) = -\hat{\mathbf{d}}_{mol} \cdot \mathbf{F}_M(t)$ and $\hat{V}_C(t) = -\hat{\mathbf{d}}_{mol} \cdot \mathbf{F}_M(t)$ (first superscript $d$ and $\bar{d}$), respectively. In eq 31b, we fully consider cavity field effects, whereas in eq 31a, we do not.

From eqs 29 and 31b, we get for the cavity field case

$$\bar{\Sigma}^{\bar{d}\bar{d}}(\omega) = \frac{4\pi\omega}{c} \frac{Im[\boldsymbol{\alpha}^{\bar{d}\bar{d}*}(\omega)\mathbf{F}_M^*(\omega)\cdot\mathbf{F}_M(\omega)]}{|\mathbf{F}_M(\omega)|^2 n(\omega)} \tag{32}$$

Performing an orientational average over all the possible field orientations, we get

$$\bar{\Sigma}_{avg}^{\bar{d}\bar{d}}(\omega) = \frac{4\omega}{c} \frac{Im\left[\frac{1}{3}\sum_{i=1}^{3} \alpha_{ii}^{\bar{d}\bar{d}}(\omega)\right]}{n(\omega)} \tag{33}$$

Analogously, for the case without cavity field effects, we have

$$\Sigma_{avg}^{dd}(\omega) = \frac{4\omega}{c} \frac{Im\left[\frac{1}{3}\sum_{i=1}^{3} \alpha_{ii}^{dd}(\omega)\right]}{n(\omega)} \tag{34}$$

The key quantities in the optical spectra of eqs 33 and 34 are the polarizability tensors $\boldsymbol{\alpha}^{dd}(\omega)$ and $\boldsymbol{\alpha}^{\bar{d}\bar{d}}(\omega)$ with and without cavity field effects, respectively. The polarizability tensor components may be written as

$$\alpha_{ij}^{\bar{d}\bar{d}}(\omega) = \frac{\bar{d}_i(\omega)}{F_{M,j}(\omega)} \tag{35a}$$

$$\alpha_{ij}^{dd}(\omega) = \frac{d_i(\omega)}{F_{M,j}(\omega)} \tag{35b}$$

Particularly, we are interested only in the diagonal terms, since such are the components relevant for a photoabsorption cross section spectrum obtained from unpolarized light (see in eqs 33 and 34). We stress that $\mathbf{d}(\omega)$ and $\bar{\mathbf{d}}(\omega)$ are induced dipole moments, not permanent ones. We can write them using the Fourier transform from the time-domain as

$$\bar{\mathbf{d}}(\omega) = \int_{-\infty}^{\infty} (\bar{\mathbf{d}}(t) - \bar{\mathbf{d}}(0))e^{i\omega t} \, dt \tag{36a}$$

$$\mathbf{d}(\omega) = \int_{-\infty}^{\infty} (\mathbf{d}(t) - \mathbf{d}(0))e^{i\omega t} \, dt \tag{36b}$$

To obtain the optical spectrum, an impulsive electric field independent of the position and with a Dirac time-dependence, that is, $\mathbf{F}_M(\mathbf{r}, t) = \mathbf{F}_M^0 \delta(t)$ is often employed as a perturbator of the system. Being frequency-independent, the latter field is suitable to perturb equally all the dipole-allowed excitations of the system. Using this kind of external (Maxwell) electric field, we get the full expression for the polarizability tensor components as

$$\alpha_{ij}^{\bar{d}\bar{d}}(\omega) = \frac{1}{F_{M,j}^0} \int_{-\infty}^{\infty} dt e^{i\omega t}(\bar{\mathbf{d}}(t) - \bar{\mathbf{d}}(0))$$

$$\alpha_{ij}^{dd}(\omega) = \frac{1}{F_{M,j}^0} \int_{-\infty}^{\infty} dt e^{i\omega t}(\mathbf{d}(t) - \mathbf{d}(0)) \tag{37}$$

The time-dependent effective and bare molecular dipole moments are written as

$$\bar{\mathbf{d}}(t) = \mathbf{d}_{mol}^{cf}(t) + \tilde{\mathbf{d}}(t) \tag{38a}$$

$$\mathbf{d}_{mol}(t) = \int \rho(\mathbf{r}, t)\mathbf{r} \, d\mathbf{r} \tag{38b}$$

where $\mathbf{d}_{mol}^{cf}(t)$ is also computed from eq 38b in the case of a cavity-field time-dependent simulation, and

$$\tilde{\mathbf{d}}(t) = -\int_{-\infty}^{\infty} \int \nu_{mol}(\mathbf{s}, t') \frac{\partial \alpha_{cf}}{\partial \mathbf{F}_M}(\mathbf{s}, t - t') \, dA(\mathbf{s}) \, dt' \tag{39a}$$

$$\nu_{mol}(\mathbf{s}, t) = \int \frac{\rho(\mathbf{r}, t)}{|\mathbf{r} - \mathbf{s}|} d\mathbf{r} \tag{39b}$$

The time-dependent convolution in eq 39a is unsuitable for an implementation due its high computational cost. Therefore, to compute this part of the cavity field contribution to the optical absorption spectrum we follow a different approach stemming from an alternative partition for the total electro-static potential (see Appendix A.1, cf., eq 1), and described in the Appendix A.2.

## 3. RESULTS

### 3.1. Computational Details.
To illustrate the operation of our methodology accounting for nonequilibrium cavity field effects within TDPCM, we perform RT-TDDFT simulations and compute the optical absorption spectrum of a case-study solvated system. We choose azobenzene in its trans configuration as a solute molecule. *trans*-Azobenzene geometry is planar, and it is optimized in vacuo at the DFT level. We work under the adiabatic local density approximation (LDA), and we use standard pseudopotentials consistent with the LDA exchange-correlation functional. The solvent is described using a Debye dielectric model with parameters $\tau_D = 3.37$ ps, $\epsilon_0 = 78.39$, and $\epsilon_d = 1.776$, mimicking water.[30,31] We perform as well linear response TDDFT (LR-TDDFT) calculations of *trans*-azobenzene in vacuo to obtain the first-excited state transition dipole moment (TDM). To that aim, all the single-particle transitions in an energy window of ~10 eV about the Fermi energy are considered. Molecule-solvent interactions via the reaction field are always taken into account, regardless of the inclusion of cavity field effects or not. To carry out all computations, we use a modified version of Octopus (version 8.0).[32] (see Appendix A.5 for details on the implementation).

In our calculation, the real-space grid is built from interlocking spheres of 5 Å of radius centered at the atomic positions, with a spacing between grid points of 0.18 Å. The real-time simulations were performed for 15 fs with a time step of 1.5 as. We start the propagation of the KS orbitals from the ground state after applying a *weak* and *impulsive* external electric field with an amplitude of ~5 × 10⁻³ a.u.

### 3.2. Optical Absorption Spectrum with Cavity Field Effects.
In this section, we present the results for the optical absorption spectrum of *trans*-azobenzene in vacuo and in aqueous solution, focusing on cavity field effects.

As anticipated, we obtain the optical absorption spectrum of the solvated molecule by applying an impulsive electric field oriented along each of the three Cartesian directions. The averaged photoabsorption cross-section is calculated using eqs 33 and 34, respectively, if we consider cavity field effects or not. $\boldsymbol{\alpha}_{\bar{d}\bar{d}}(\omega)$ and $\boldsymbol{\alpha}_{dd}(\omega)$ are computed from the Fourier transform of the time-dependent effective or bare dipole moments (eq 37). Finally, we compute the time-dependent dipoles via the time-dependent density and reaction-field polarization charges (eqs 38b, 53, and 58) obtained from RT-

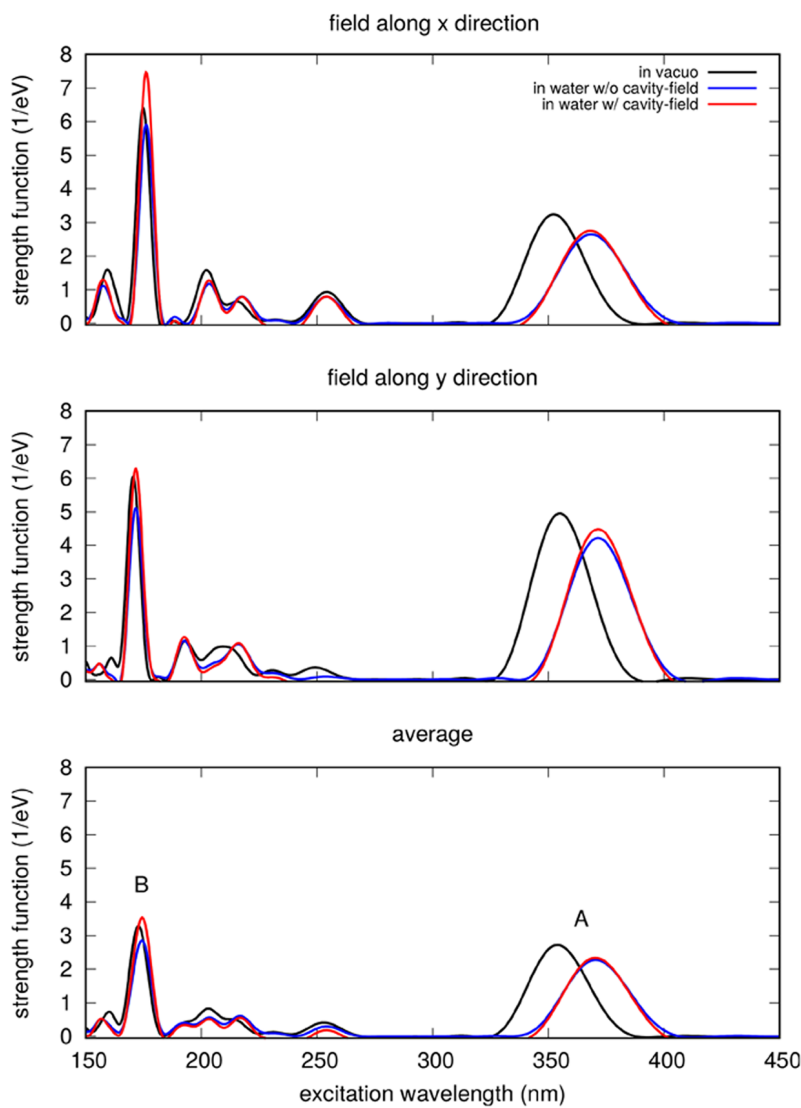

**Figure 1.** Absorption spectra of *trans*-azobenzene in vacuum (black lines) and water. The case with (without) cavity field effects is represented with red (blue) lines. For the first two panels, the impulsive electric field is parallel to the plane of the molecule (see Figure 2). The bottom panel shows the absorption spectrum averaged over the three Cartesian directions of the field (eqs 33 and 34).

TDDFT calculations coupled with EOM-PCM, as described in ref 13 and section 2.2 for the reaction and cavity field solvent polarization, respectively.

In Figure 1, we show the results for the optical absorption spectra (strength function ∝ cross-section) for the molecule in vacuo and in solution, with and without cavity field effects. The first two panels contain the absorption spectra using an electric field in the plane of the molecule (see Figure 2). In the last panel, we plot the mean value of the absorption spectra in the three Cartesian directions.

In addition to the RT-TDDFT computations, we perform a LR-TDDFT calculation in vacuo to obtain the TDMs corresponding to the selected transitions A an B indicated in the lower panel of Figure 1. The TDM associated with the first optically active excited state (namely, transition A), can be directly computed from LR-TDDFT results, using the appropriate transition density, as $d_{0A} = \int \rho_{0A}(r)r \, dr$. (We drop the subscript "mol" for the sake of simplicity.) However, LR-TDDFT and RT-TDDFT spectra differ substantially in the high-frequency region, making difficult to match the absorption peaks and hence the underlying excitations. Therefore, to get the TDM associated with the high-energy transition B, $d_{0B}$, we follow a different path. To find the magnitude of $d_{0B}$, we use the fact that ratio between absorption peak strengths is

$$f_{AB,\alpha} = \frac{|d_{0B,\alpha}|^2 \Delta E_{0B}}{|d_{0A,\alpha}|^2 \Delta E_{0A}} \quad (40)$$

where $\Delta E_{0A}$ and $\Delta E_{0B}$ are the excitation energies corresponding to excitations A and B, respectively, and $\alpha$ indicate Cartesian components $x$, $y$, $z$. We extract $\Delta E_{0A}$ and $\Delta E_{0B}$ from the average absorption spectrum (see Figure 1 bottom panel), whereas $f_{AB,\alpha}$ is taken from the absorption spectra with the field along $\alpha$ direction (see Figure 1 top and middle panel for $\alpha = x$, $y$); both quantities are obtained from RT-TDDFT calculations. Using, in addition, the LR-TDDFT result for $d_{0A}$, we obtain | $d_{0B}$|. To get the direction of $d_{0B}$, that is, $d_{0B}/|d_{0B}|$, we use RT-TDDFT results directly. We extract the optical absorption cross section matrix $\Sigma_{ij}$ at the energy of the transition B, and we diagonalize it. For a nondegenerate band, we expect just one eigenvalue different from zero, and since $\Sigma_{ij} = \frac{4\pi\omega}{c} d_{0B}^i d_{0B}^j$,[21,33] the normalized eigenvector associated with the nonzero eigenvalue would set the direction for the TDM, $\mathbf{d}_{0B}$. If in turn the band is almost degenerate, we can still find the TDMs with the largest contributions to the absorption by diagonalizing the cross-section matrix and taking the eigenvectors associated with the highest eigenvalues. In our case, we find at least two possible directions for the B band TDMs that we will call $B_1$ and $B_2$ transitions. Results for A, $B_1$, and $B_2$ TDMs are reported in Table 1. Figure 2 shows a schematics of *trans*-azobenzene molecule along with the TDMs corresponding to A, $B_1$, and $B_2$ excited states.

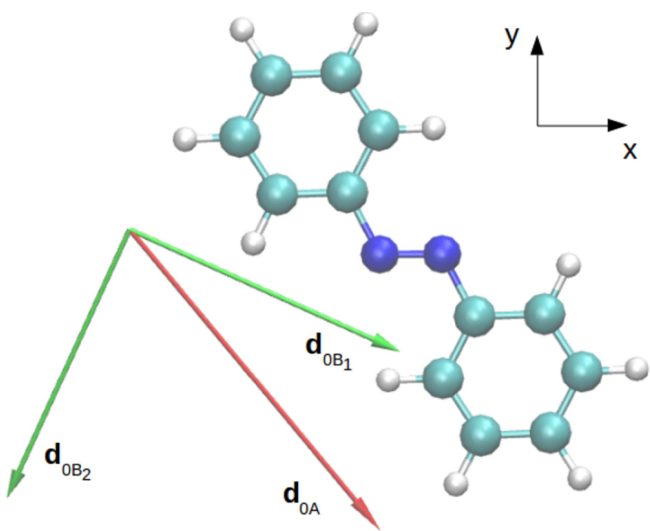

**Figure 2.** Optimized structure of *trans*-azobenzene in vacuo, along with TDMs associated with A and B bands from Table 1.

**Table 1. TDMs Associated with A and B Bands in Figure 1[a]**

|        | $d_{0A}$ (D) | $d_{0B_1}$ (D) | $d_{0B_2}$ (D) |
|--------|--------------|----------------|----------------|
| $x$    | 5.02         | 5.42           | −2.46          |
| $y$    | −6.07        | −2.46          | −5.42          |
| $z$    | 0.00         | 0.00           | 0.00           |

[a]Since the B band is degenerate, results are shown for two excited states, $B_1$ and $B_2$, contributing the most to B absorption.

Returning to Figure 1, first, we observe a red shift in the absorption peaks of the solvated molecule with respect to its gas-phase counterparts. The latter is an expected reaction field effect. Cavity−field effects might induce shifts in the spectrum when the solvent dielectric function dispersion in the absorption frequency region is large. In this case, the red and blue ends of an absorption band are affected differently by the cavity field effect, and the band would be deformed. Such a strong dispersion is however not the case for common solvent, and we do not expect shifts associated with cavity field effects.

Second, we check that the cavity field effects change not only the absolute but also the relative strength of absorption peaks, therefore impacting in a nontrivial way on the absorption spectrum. To assess cavity field effects for *trans*-azobenzene in solution in a quantitative way, we compare our results for the cavity field factor (CFF) with the prediction by Onsager model, which considers the molecule as hosted by a spherical cavity.[1] We select two representative optically active transitions (called A and B in Figure 1) and we compute the CFF directly

from the spectra for transitions A and B (we refer to it as "PCM" CFF or just CFF), using

$$\text{CFF} = \frac{\overline{\Sigma}_i^{\overline{d}\overline{d}}(\omega)}{\Sigma_i^{dd}(\omega)} \tag{41}$$

and through the Onsager analytic formula, that is,

$$\text{Onsager CFF} = \left| \frac{3\epsilon(\omega)}{2\epsilon(\omega) + 1} \right|^2 \tag{42}$$

The results are reported in Table 2. Notice that not only there is a substantial quantitative departure from Onsager prediction for both A and B transitions, but also there is a strong dependence of the CFF on the specific transition that is overlooked by Onsager model.

**Table 2. CFF Computed from Eq 41 Compared with the Prediction by Onsager Model, Eq 42, for Two Selected Transitions from Figure 1**

| transition | wavelength (nm) | CFF  | Onsager CFF |
|------------|-----------------|------|-------------|
| A          | 369             | 1.03 | 1.37        |
| B          | 174             | 1.24 | 1.37        |

Since the Debye dielectric function for water is almost constant in the ultraviolet and visible (UV−vis) spectrum, neither the PCM (eq 41) nor Onsager CFF (eq 42) are able to bear any frequency dependence through the dielectric constant. The reason for the CFF difference for transitions A and B is instead explored in the next section.

*Anistropic Cavity Field Effects.* Keeping the dielectric function as a constant, the ultimate cause for the disagreement between PCM and Onsager model comes from the fact that the solvent−molecule interface cannot be adequately modeled by a spherical surface for the molecule in question (see Figure 2). In the Onsager (spherical) model, the electric field acting on the molecule (cavity field) is increased by the same amount independently on the polarization direction and the frequency. Therefore, cavity field effects treated in such fashion, provide equal enhancement of the strength of every peak in the average absorption spectrum (cf. lower panel in Figure 1). If we consider more realistic shapes for the molecular cavity, the absorption strength corresponding to certain excitations would be significantly more enhanced than others depending on the orientation of their TDMs. Even in the case of transitions that are associated with TDMs that are practically colinear this effect may be substantial.

To illustrate the latter, let us consider a still simple but more physically informed shape for the molecular cavity. We remark that *trans*-azobenzene is more than 2.5 larger in the direction connecting the two benzene moieties than in the orthogonal direction. We notice as well that the transitions of interest are characterized by in-plane dipole moments (see Figure 2). To describe the cavity field influence in these two transitions, we are merely interested in addressing the possible effect of the cavity anisotropy in the plane. Hence, we need a cavity featuring different characteristic length in the "long" and "short" axes of the planar structure, disregarding the chosen (perhaps arbitrary) characteristic length in the third direction orthogonal to the molecular plane. We take advantage of the known analytic expression for the CFF in the case a prolate ellipsoidal cavity.[14,34,35] We choose the prolate ellipsoid with the major axis along the direction of the TDM corresponding to A excited state, that is, $d_{0A}$, and having a circumference section with a diameter 0.38 times as long as the major axis of the in-plane ellipse. Then, we calculate the CFF as the squared ratio between TDMs with and without cavity field effects.[14] The computation of TDM with cavity field effects within the prolate ellipsoidal model amplifies more the components in the shorter axis.[14,35] This setup is sufficient to explain the transition-dependence of the CFF, yielding a CFF = 1.12 for A, 1.20 and 1.46 for $B_1$ and $B_2$, respectively. From the last numbers, we recover the correct trend of cavity field effects, that is, the largest CFF is obtained for the excited states with largest component of the TDM along the shorter axis of the molecule, namely, $B_2$ transition in our case. Within the ellipsoidal model, we have an overall CFF = 1.33 for the B band (assuming $B_1$ and $B_2$ TDMs to be exactly orthogonal) which match qualitatively the PCM data reported in Table 2. We do not expect to find a quantitative agreement through the former analysis since the molecular cavity is in any case more complex than an ellipsoid and better described within PCM.

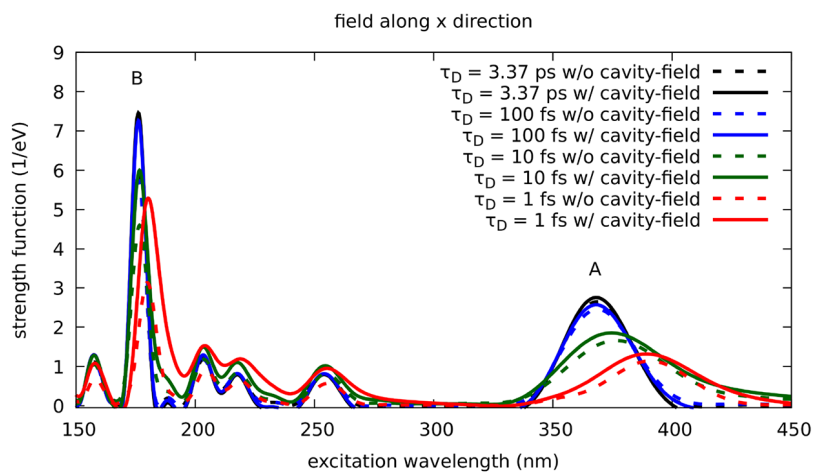

**Figure 3.** Absorption spectra of *trans*-azobenzene in solution featuring cavity field effects or not for solvents with different relaxation times.

*Nonequilibrium Cavity Field Effects.* Finally, we comment on the nonequilibrium effects as treated using EOM-TDPCM. In section 2.2, we obtained the expression for the polarization charges propagation in the case of infinite relaxation time, that is, eq 13. Using this limit in both the reaction-field and cavity-field time-dependent polarization charges, we obtained the same absorption spectrum as for the EOM-TDPCM plotted in Figure 1 (not shown). The reason is that the Debye relaxation of water is much longer ($\tau_D = 3.37$ ps) than the characteristic time for electronic excitations (tenth of fs). The slow degrees of freedom of the solvent (driving the inertial polarization) cannot follow the rapid change of either the electronic density or the Maxwell field in the frequency range of electronic excitations (the UV–vis spectrum). Therefore, the dynamics of the solvent polarization relaxation after the application of the impulsive field is dominated only by its fast degrees of freedom, which is characterized by the high-frequency dielectric constant of the solvent, $\epsilon_d$ (see eq 13). As a consequence, the only transition-dependence appearing in the CFF is coming from the anisotropic cavity, and not from the dielectric function frequency dependence, which is the ultimate responsible for nonequilibrium solvent polarization effects. Nevertheless, this situation might be drastically different when the relaxation time of the solvent is small enough to be comparable with the characteristic time of electronic excitations. For these cases, the strong frequency dependence of the dielectric function in the electronic excitation range reflects on the CFF. Such a CFF frequency dependence is a unique feature of nonequilibrium cavity field effects in the absorption spectrum that is unrelated to the anisotropy of the cavity.

To expose nonequilibrium cavity field effects, we compute the CFF for the two selected transitions (namely, A and B) for a set relaxation times and ascertain that (i) the CFF ($CFF_B$ and $CFF_A$) and (ii) the CFF difference ($CFF_B - CFF_A$) change with the relaxation time. This dependence shows that there is an underlying frequency dependence in the CFF that cannot be accounted for by considering anisotropic cavity field effects.

Therefore, we performed further EOM-TDPCM/RT-TDDFT calculation on the solvated molecule for hypothetical solvents with shorter Debye relaxation times and with the same static and dynamic dielectric constant of water. For the sake of simplicity, we choose only one direction for the application of the impulsive field, namely, the x-axis (see Figure 2). The results for the optical absorption spectra of solvents with $\tau_D$ = 3370 (water), 100, 10, and 1 fs are plotted in Figure 3 for comparison.

In Figure 3, we observe a great variation in the spectra featuring no cavity field effects for different relaxation times. In particular, we notice a systematic redshift of the first absorption peak (transition A) for diminishing relaxation times. This strong effect is due to reaction field and its discussion is out of the scope of this Article. We rather focus on the variation of CFF with the solvent relaxation time in the case of transitions A and B. A summary of these results is reported in Table 3.

In Table 3, we observe a clear CFF relaxation-time dependence for both transitions, our first evidence that nonequilibrium cavity field effects are taken into account. In particular, we detect a monotonic behavior of the CFF with the diminishing relaxation time that can be explained qualitatively on the grounds of Onsager model. For any given frequency, Onsager CFF increases toward a maximum value for small relaxation times, that is, CFF $\xrightarrow{\tau_D \to 0}$ $|3\epsilon_0/(2\epsilon_0 + 1)|^2 = 2.22$, and decreases toward a minimum value for large relaxation times, that is, CFF $\xrightarrow{\tau_D \to \infty}$ $|3\epsilon_d/(2\epsilon_d + 1)|^2 = 1.37$. The ellipsoidal cavity model (vide supra) does not change the qualitative picture drawn from Onsager's model for CFF versus $\tau_D$; it simply tunes the maximum and minimum CFF at zero or infinite relaxation times and has excitation-dependent minimum and maximum values. The PCM, featuring a more complex molecular-shape cavity, further adjust quantitatively these maximum and minimum CFF values (as seen in Table 3). In each model, the maximum and minimum CFF values for zero or infinite relaxation times are associated with the static and dynamic dielectric constants of the solvent, respectively. For the spherical model, the minimum and maximum CFF are independent from the frequency, whereas in the ellipsoidal model they depend on the orientation of the underlying transition dipole moment (anisotropic cavity effects). In the last row of Table 3, we show the difference $CFF_B - CFF_A$. As expected, the difference in CFF is not constant, which is another evidence of CFF dependence on the frequency unaccounted by anisotropic cavity effects. Therefore, nonequilibrium cavity field effects have an impact in the absorption spectral shape.

**Table 3. CFF Computed from Eq 41 for Transitions A and B for Solvents with Different Relaxation Times[a]**

| $\tau_D$ (fs) | 3370 | 1000 | 100 | 10 | 1 |
|---|---|---|---|---|---|
| $CFF_A$ | 1.04 | 1.04 | 1.05 | 1.12 | 1.16 |
| $CFF_B$ | 1.26 | 1.26 | 1.26 | 1.30 | 1.70 |
| $CFF_B - CFF_A$ | 0.22 | 0.22 | 0.21 | 0.18 | 0.54 |

[a]The case of 3.37 ps is the water-like solution.

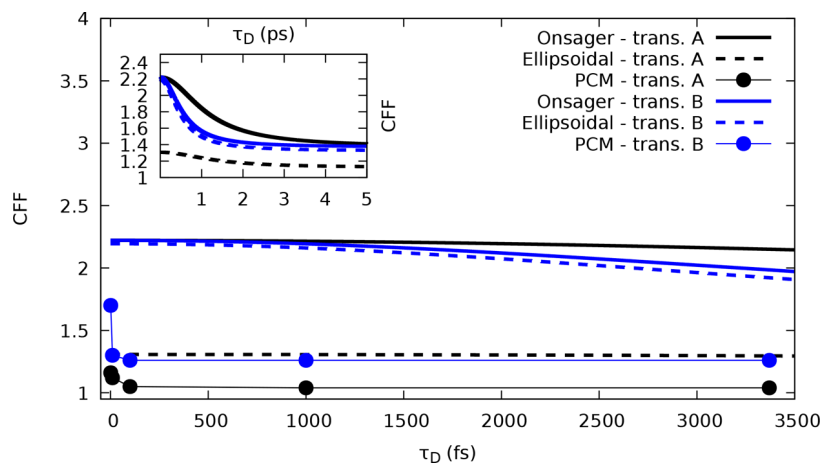

**Figure 4.** CFF corresponding to transitions A and B of *trans*-azobenzene in solution vs the solvent relaxation time $\tau_D$. Solid and dashed lines represent the spherical (Onsager) and prolate ellipsoidal cavity models, respectively, whereas points are associated with PCM results with a molecular-shape cavity. PCM data points are joined as a guide for the eye. Black (blue) color refers to A (B) transition. The inset shows the spherical and ellipsoidal models curves in a different scale.

In Figure 4, we plot the CFF versus $\tau_D$ for the spherical (Onsager), the prolate ellipsoidal and the molecular-shape (PCM) cavity models. For the ellipsoidal model, we plot the overall CFF corresponding to the B band (before decomposed in $B_1$ and $B_2$), to facilitate its comparison with PCM results. Apart from the quantitative departure of the molecular-shape model from the rest, there is an important qualitative difference: the transient between maximum to minimum CFF values is narrower, occurring in a relaxation-time window of tenth of femtosecond for the molecular-shape cavity, as opposed to a few picoseconds for spherical and ellipsoidal cavities (see Figure 4 inset). The spherical and ellipsoidal cavity models do not predict relevant nonequilibrium cavity field effects for $\tau_D \leq 0.5$ ps or $\tau_D \geq 3$ ps, where the CFF takes approximately its maximum and minimum values, respectively. We recall that the latter might be transition-dependent for the ellipsoidal model because of the cavity anisotropy. It is in the relaxation time window 0.5−3 ps, where we expect to find nontrivial nonequilibrium cavity field effects in the form of strongly frequency-dependent CFF. However, in the case of a molecular-shape cavity, we find that nonequilibrium cavity field effects can only be relevant in the relaxation time scale of 0−100 fs. The latter disagreement highlights the fact that an accurate description of the molecular cavity is also essential to account for nonequilibrium cavity field effects, making impossible to decide when to apply our full EOM-TDPCM treatment (with cavity field effects) based on estimations extracted from simple models (such as Onsager's).

## 4. CONCLUDING REMARKS

In summary, we have extended the time-dependent polarizable continuum model (TDPCM) framework,[2,5] in its equation-of-motion (EOM-TDPCM) version,[13] to account for nonequilibrium cavity field effects within quantum mechanical calculations of molecules in solution and subject to external electrostatic fields and/or electromagnetic radiation (light). Such a scheme (EOM-TDPCM with cavity field effects) provides a more realistic picture for the description of solvated molecule, shortening the gap between simulation/theoretical and experimental results.

Our theoretical contribution has been incorporated in the Octopus code (to be released as version 9.0).[32] This implementation enables us to perform real-time time-dependent density functional theory (RT-TDDFT) simulations coupled with EOM-TDPCM and optical absorption spectrum calculations, including nonequilibrium reaction and cavity field effects.

To show an application of our methodology, we compute the optical absorption spectrum of *trans*-azobenzene in water employing EOM-TDPCM/RT-TDDFT simulations. These calculations are relevant to highlight the nontrivial role of anisotropic and nonequilibrium cavity field effects in the optical absorption spectrum of solvated molecules, however, they are not intended to provide benchmarking results for *trans*-azobenzene in water. We underline that the cavity field factor, that is, the ratio between the absorption spectra with and without cavity field effects, changes throughout the UV−vis. spectrum frequencies depending strongly on the orientation of the TDMs of the excited states associated with each absorption peak.

Our cavity field effects description within PCM goes beyond the simplified Onsager's model, where the molecule is hosted by a spherical cavity.[3,4] At difference with Onsager's model, PCM give rise to a modification of the optical absorption intensities that is transition-dependent. This effect was rationalized in terms of another simplified model for an anisotropic cavity with ellipsoidal shape, which is able to reproduce qualitatively the PCM results.

Finally, we show that nonequilibrium solvent polarization affects the overall shape of the optical absorption spectrum. In particular, we find evidence of the nonequilibrium effects in the change of absorption peak height due to the cavity field for different relaxation times. We remark that an accurate description of the cavity shape is so important to our treatment of nonequilibrium cavity field effects that we cannot even rely on the typical Onsager's (spherical cavity) model to conclude whether a nonequilibrium treatment is required or not.

## ■ APPENDIX

### A.1. Alternative Formulation of Static Polarization

The partition in eq 1 is arbitrary. We may as well pose it in a different way, akin to the standard PCM formulation for the reaction-field response, namely

$$v(\mathbf{r}) = f(\mathbf{r}) + v_{cf}'(\mathbf{r}) \tag{43}$$

We take $v_C(\mathbf{r}) \equiv v(\mathbf{r})$ for $\mathbf{r} \in \Omega_i$. Inside the cavity, $f(\mathbf{r})$ is equal to the potential generated by the fictitious external charges (source of the Maxwell field) in a homogeneous vacuum. Elsewhere it matches the potential associated with the Maxwell field (applied in a dielectric media without the vacuum cavity). In the most general case, $f(\mathbf{r})$ can be represented in terms of the Green functions for the Poisson problems in and outside the cavity, respectively. For a solvent, this expression reduces to

$$f(\mathbf{r}) = \begin{cases} \epsilon v_M(\mathbf{r}) & \text{if } \mathbf{r} \in \Omega_i \\ v_M(\mathbf{r}) & \text{if } \mathbf{r} \in \Omega_e \end{cases} \tag{44}$$

As a consequence of this partition, the potential $f(\mathbf{r})$ is discontinuous at $\mathbf{r} \in \Gamma$, whereas the normal component of the electric displacement vectors corresponding to $f(\mathbf{r})$ is instead continuous across the interface $\Gamma$. The term $v_{cf}'(\mathbf{r})$ is again a polarization contribution that solves the Laplace problem in and outside the cavity, as its counterpart $v_{cf}(\mathbf{r})$ in the partition eq 1 (see section 2.1), but this time obeying a different set of continuity conditions able to recover those from the original generalized Poisson problem.

In this case, eq 4 holds with a different PCM response matrix, namely

$$Q_{cf}' = -[\mathbb{S}_e(2\pi\mathbb{1} + \mathbb{D}_i^*\mathbb{A}) + (2\pi\mathbb{1} - \mathbb{D}_e\mathbb{A})\mathbb{S}_i]^{-1} \\ \times [(2\pi\mathbb{1} - \mathbb{D}_e\mathbb{A})(\mathbb{S}_i\mathbb{S}_e^{-1} - \mathbb{1})] \tag{45}$$

which particularizing to the solvent case, yields

$$Q_{cf}' = -\mathbb{S}_i^{-1}\left(2\pi\frac{(\epsilon+1)}{(\epsilon-1)}\mathbb{1} - \mathbb{D}_i\mathbb{A}\right)^{-1}(2\pi\mathbb{1} - \mathbb{D}_i\mathbb{A})\epsilon \tag{46}$$

Hence, within partition eq 43, the cavity-field PCM response matrix (eq 7) in the latter case distinguishes from its reaction-field counterpart only by an extra factor, $\epsilon$. The extra factor indicates that the effect of the Maxwell potential should be taken inside the vacuum cavity since the polarization potential of eq 3 is the one felt by a distribution of charges placed therein.

### A.2. Alternative Formulation to the Optical Spectrum

Let us find an expression for the absorption cross-section employing the partition (eq 43) instead of eq 1 in eq 24. In the case of a solvent dielectric medium, we can write explicitly $v_C(\mathbf{r},\omega) = \epsilon(\omega)v_M(\mathbf{r},\omega) + v_{cf}'(\mathbf{r},\omega)$, where $v_M(\mathbf{r},\omega)$ and $v_{cf}(\mathbf{r},\omega)$ are the Maxwell-field and polarization potentials, respectively. Using as well the quasi-static approximation for the Maxwell potential in eq 24, we get

$$\overline{\Sigma}(\omega) = \Sigma_M(\omega) + \Sigma_{cf}(\omega)$$

$$= -\frac{4\pi\omega}{c}\frac{Im[\epsilon(\omega)\mathbf{d}_{mol}^{cf*}(\omega)\cdot\mathbf{F}_M(\omega)]}{|\mathbf{F}_M(\omega)|^2 n(\omega)}$$

$$+ \frac{4\pi\omega}{c}\frac{Im\left[\int_{\mathbb{R}^3}\rho_{mol}^*(\mathbf{r},\omega)v_{cf}'(\mathbf{r},\omega)\,d\mathbf{r}\right]}{|\mathbf{F}_M(\omega)|^2 n(\omega)} \tag{47}$$

Again, $\mathbf{d}_{mol}^{cf}(\omega)$ includes implicitly cavity field effects, but it is not the only source of them in the cross-section spectrum. We can write the second term as in eq 26. Writing $v_{cf}'(\mathbf{r},\omega)$ in terms of its sources $\sigma_{cf}'(\mathbf{r},\omega)$ (N.B. $\sigma_{cf}' \neq \sigma_{cf}$), we have

$$\Sigma_{cf}(\omega) = \frac{4\pi\omega}{c}\frac{Im\left[\int_{\Gamma}v_{mol}^*(\mathbf{s},\omega)\sigma_{cf}'(\mathbf{s},\omega)\,dA(\mathbf{s})\right]}{|\mathbf{F}_M(\omega)|^2 n(\omega)} \tag{48}$$

Now, thanks to the PCM response equation from eq 4 for the external charges in integral form, eq 26 transforms into

$$\Sigma_{cf}(\omega) = \frac{4\pi\omega}{c}\frac{1}{|\mathbf{F}_M(\omega)|^2 n(\omega)} \times Im\left[\int_{\Gamma}\int_{\Gamma}v_{mol}^*(\mathbf{s},\omega)\right.$$

$$\left. Q_{cf}'(\mathbf{s},\mathbf{s}',\omega)v_M(\mathbf{s}',\omega)\,dA(\mathbf{s})\,dA(\mathbf{s}')\right] \tag{49}$$

$Q_{cf}'(\mathbf{s},\mathbf{s}',\omega)$ is the PCM response operator generating the polarization charge density due to the applied potential, that is, $\sigma_{cf}'(\mathbf{s}) = \int_{\Gamma}Q_{cf}'(\mathbf{s},\mathbf{s}',\omega)v_M(\mathbf{s}',\omega)\,dA(\mathbf{s}')$. Within partition 43, $Q_{cf}'(\mathbf{s},\mathbf{s}',\omega) = \epsilon(\omega)Q(\mathbf{s},\mathbf{s}',\omega)$, where $Q(\mathbf{s},\mathbf{s}',\omega)$ is the PCM response operator generating the polarization charge density because of the solute molecule (see eq 45), that is, $\sigma(\mathbf{s}) = \int_{\Gamma}Q(\mathbf{s},\mathbf{s}',\omega)v_{mol}(\mathbf{r},\omega)\,dA(\mathbf{s}')$. $Q(\mathbf{s},\mathbf{s}',\omega) = Q(\mathbf{s}',\mathbf{s},\omega)$ as a result of the symmetry properties of the Green's functions in the integral operator. Therefore, we can write eq 49 as

$$\Sigma_{cf}(\omega) = \frac{4\pi\omega}{c}\frac{Im\left[\epsilon(\omega)\int_{\Gamma}\sigma^*(\mathbf{s}',\omega)v_M(\mathbf{s}',\omega)\,dA(\mathbf{s}')\right]}{|\mathbf{F}_M(\omega)|^2 n(\omega)} \tag{50}$$

and using the quasi-static expression for the potential associated with the Maxwell field, we get

$$\Sigma_{cf}(\omega) = -\frac{4\pi\omega}{c}\frac{Im[\epsilon(\omega)\mathbf{d}_{cav}^*(\omega)\cdot\mathbf{F}_M(\omega)]}{|\mathbf{F}_M(\omega)|^2 n(\omega)} \tag{51}$$

where $\mathbf{d}_{cav}(\omega)$ is the dipole of the cavity polarization charges induced by the solute molecule (N.B. not by the Maxwell field). Using eq 51, the complete cross-section yields

$$\overline{\Sigma}(\omega) = -\frac{4\pi\omega}{c} \frac{Im[\overline{\mathbf{d}}^*_{mol}(\omega) \cdot \mathbf{F}_M(\omega)]}{|\mathbf{F}_M(\omega)|^2 n(\omega)} \tag{52}$$

where

$$\overline{\mathbf{d}}_{mol}(\omega) = \epsilon(\omega)(\mathbf{d}^{cf}_{mol}(\omega) + \mathbf{d}_{cav}(\omega)) \tag{53}$$

the effective induced dipole of the solvated molecule (external dipole in Onsager's notation[1]).

Equation 51 is perhaps the simplest, clearer in meaning and easy-to-code expression for calculating the cross-section due to the polarization contribution to the total applied field, compare eqs 47 and 48, but it might be numerically inconvenient.

To illustrate the latter we resort to Onsager's model,[1] where a point-dipole inserted in the center of a spherical empty cavity interacts with the external dielectric medium and with an applied electric field. In such a case, the dipole associated with the reaction-field is

$$\mathbf{d}_{cav}(\omega) = -\frac{2(\epsilon(\omega) - 1)}{2\epsilon(\omega) + 1} \mathbf{d}^{cf}_{mol}(\omega) \tag{54}$$

Hence, $\overline{\mathbf{d}}_{mol}(\omega)$ is very near zero when $\epsilon(\omega)$ is large. For instance, if we consider an equilibrium propagation, that is, $\epsilon(\omega) \equiv \epsilon_0$ with $\epsilon_0 = 100$, we get $\overline{\mathbf{d}}_{mol}(\omega) \approx \mathbf{d}^{cf}_{mol}(\omega)$ .

There is a source of numerical error coming from the subtraction of quantities with different levels of approximation, namely, $\mathbf{d}^{cf}_{mol}(\omega)$ and $\mathbf{d}_{cav}(\omega)$. $\mathbf{d}^{cf}_{mol}(\omega) = \int_{\mathbb{R}^3} \rho(\mathbf{r}, \omega)\mathbf{r} \, d\mathbf{r}$ is the exact expression used to compute the former, whereas $\mathbf{d}_{cav}(\omega) = \sum_{j=1}^{N_{tess}} q^i s_j$ is the BEM expression to compute the latter. Actually, since we use a real-space code (Octopus) the previous integral is approximated by a sum over grid points, but the accuracy of the approximation for $\mathbf{d}^{cf}_{mol}(\omega)$ and $\mathbf{d}_{cav}(\omega)$ is different, being cruder the one for $\mathbf{d}_{cav}(\omega)$ . The combination of the two approximations, plus the amplification of the errors due to the product by $\epsilon(\omega)$ in eq 29 (a fortiori, when the dielectric function values are large) implies that in some cases eq 51 might produce erroneous results. Equations 47 or 48 might be of use when numerical errors hinder the application of eq 51.

Alternatively, to account for the full cavity field effects in the cross-section spectrum of eq 33 we may still circumvent the calculation of the effective induced dipole ($\overline{\mathbf{d}}_{mol}$) exploiting the ratio between the cross-section spectra coming from the induced molecular dipole with ($\mathbf{d}^{cf}_{mol}$) and without cavity field effects ($\mathbf{d}_{mol}$). It yields,

$$\overline{\Sigma}^{\overline{dd}}_{avg}(\omega) = \Sigma^{dd}_{avg}(\omega) \times \left( \frac{\Sigma^{\overline{dd}}_{avg}(\omega)}{\Sigma^{dd}_{avg}(\omega)} \right)^2 \tag{55}$$

where $\Sigma^{\overline{dd}}_{avg}(\omega) = \overline{\Sigma}^{\overline{dd}}_{avg}(\omega) - \Sigma_{cf}(\omega)$. $\Sigma^{dd}_{avg}(\omega)$ is written in eq 34. The term between brackets is the frequency-dependent cavity field factor for the photoabsorption cross-section spectrum, that is, the ratio between the cross-section with and without cavity field effects, while monitoring the same observable, that is, the induced molecular dipole.

### A.3. Connection Formulas between the Alternative Approaches

Here, we derive some connection formulas between the two alternative partitions for the total electrostatic potential in the solvent case. Both, the effective and bare dipole moment $\overline{\mathbf{d}}_{mol}$ and $\mathbf{d}_{mol}$ are independent of the electrostatic potential partition (eqs 1 and 43). Therefore, we have

$$\mathbf{d}^{cf}_{mol}(\omega) + \tilde{\mathbf{d}}(\omega) = \epsilon(\omega)(\mathbf{d}^{cf}_{mol}(\omega) + \mathbf{d}_{cav}(\omega)) \tag{56}$$

which is valid for any cavity shape.

We recall the definition of $\mathbf{d}_{cav}$.

$$\mathbf{d}_{cav}(\omega) = \int_\Gamma \mathbf{s}\sigma(\mathbf{s}, \omega)dA(\mathbf{s}) \tag{57}$$

where

$$\sigma(\mathbf{s}, \omega) = \int_\Gamma Q(\mathbf{s}, \mathbf{s}', \omega)v_{mol}(\mathbf{s}', \omega) \, dA(\mathbf{s}') \tag{58}$$

By using the symmetry properties of the Green's function and the quasi-static approximation for the Maxwell potential, we may rewrite eq 28 as

$$\tilde{\mathbf{d}}(\omega) = \int_\Gamma \mathbf{s}\tilde{\sigma}(\mathbf{s}, \omega) \, dA(\mathbf{s}) \tag{59}$$

where

$$\tilde{\sigma}(\mathbf{s}, \omega) = \int_\Gamma Q_{cf}(\mathbf{s}, \mathbf{s}', \omega)v_{mol}(\mathbf{s}', \omega) \, dA(\mathbf{s}') \tag{60}$$

Notice the parallel between eqs 57 and 59, as well as between eqs 58 and 60.

For a spherical cavity and within Onsager's (point−dipole) model, we have[13]

$$\mathbb{D}_i\mathbf{v}_{mol}(\omega) = -\frac{2\pi}{3}\mathbf{v}_{mol}(\omega) \tag{61}$$

Using the latter and the definition of $\mathbb{Q}$ and $\mathbb{Q}_{cf}$ (see eq 7 and ref 13), we get

$$\tilde{\sigma}(\mathbf{s}, \omega) = -\frac{1}{2}\sigma(\mathbf{s}, \omega) \tag{62}$$

$$\tilde{\mathbf{d}}(\omega) = -\frac{1}{2}\mathbf{d}_{cav}(\omega) \tag{63}$$

By combining eqs 56 and 63, we obtain eq 54 and the Onsager CFF in eq 42.

### A.4. Impulsive Maxwell Potential

In this Appendix, we find the initial conditions for the propagation of polarization charges induced by an instantaneous electrostatic perturbation, that is,

$$v_M(\mathbf{r}, t) = v_{M,0}(\mathbf{r})\delta(t) \tag{64}$$

The latter is usually employed to obtain the optical spectrum of molecules from real-time density propagation. This form of the Maxwell potential excites all oscillation modes at once without changing the initial density of the system.[21] In the long wavelength limit, a uniform electric field can be used, that is, $v_{M,0} = -\mathbf{K}\cdot\mathbf{r}$.

In the frequency domain, polarization charges due to the impulse are given by

$$\mathbf{q}_{cf}(\omega) = \mathbb{Q}_{cf}(\omega)\mathbf{v}_{M,0} \tag{65}$$

where $\mathbf{v}_{M,0} = \{-\mathbf{K}\cdot\mathbf{s}_j\}$, being $\mathbf{s}_j$ a point in tessellation of $\Gamma$.

The Dirac delta time-dependence of the field is inconvenient for numerical real-time propagation schemes—a difficulty that can be easily overcome by starting the propagation at the first time-step with suitable initial conditions. In the following, we obtain such initial conditions for $\mathbf{q}_{cf}(t)$.

Plugging eq 64 into the analogous to eq 30 in ref 13 for the cavity field, we get

$$\mathbf{q}_{\mathrm{cf}}(t) = \mathbb{Q}_{\mathrm{cf,d}}\mathbf{v}_{\mathrm{M}}(t) + \Delta\mathbf{q}_{\mathrm{cf}}(t) \tag{66}$$

The first term in eq 66, gives rise to an "impulsive" polarization potential, that is, having a Dirac delta time-dependence. To deal with this term, we may use the same strategy used for the impulsive Maxwell potential.[21] Using the explicit time-dependence of eq 64 and

$$v_{\mathrm{cf,d}}(\mathbf{r}, t) = v_{\mathrm{cf,d,0}}(\mathbf{r})\delta(t)$$

$$v_{\mathrm{cf,d,0}}(\mathbf{r}) = \sum_{j=1}^{N_{\mathrm{teg}}} \frac{q_{\mathrm{cf,d,0}}^{j}}{|\mathbf{r} - \mathbf{s}_j|}$$

$$\mathbf{q}_{\mathrm{cf,d,0}} = \mathbb{Q}_{\mathrm{cf,d}}\mathbf{v}_{\mathrm{M,0}} \tag{67}$$

the initial condition for the KS orbitals is written as

$$\psi_i(\mathbf{r}, 0^+) = e^{i(\mathbf{K}\cdot\mathbf{r} + v_{\mathrm{cf,d,0}}(\mathbf{r}))}\psi_i(\mathbf{r}, 0) \tag{68}$$

The second term in eq 66 is the solution of the EOM

$$\Delta\dot{\mathbf{q}}_{\mathrm{cf}}(t) = -\mathbb{R}\Delta\mathbf{q}_{\mathrm{cf}}(t) \tag{69}$$

with the initial condition

$$\Delta\mathbf{q}_{\mathrm{cf}}(0^+) = (\tilde{\mathbb{Q}}_{\mathrm{cf,0}} - \tilde{\mathbb{Q}}_{\mathrm{cf,d}})\mathbf{v}_{\mathrm{M,0}} \tag{70}$$

The solution of this EOM is

$$\Delta\mathbf{q}_{\mathrm{cf}}(t) = \mathbb{Q}_{\mathrm{cf,D}}(t)\mathbf{v}_{\mathrm{M,0}}$$

$$\mathbb{Q}_{\mathrm{cf,D}}(t) = -\mathbb{S}^{-1/2}\mathbb{K}_{\mathrm{D}}(t)\mathbb{T}^{\dagger}\mathbb{S}^{-1/2}$$

$$\mathbb{K}_{\mathrm{cf,D}}^{ii}(t) = (K_{\mathrm{cf,0}}^{ii} - K_{\mathrm{cf,d}}^{ii})\frac{1}{\tau_{ii}}e^{-t/\tau_{ii}}\Theta(t) \tag{71}$$

Let us analyze some limiting cases. Notice that when the external medium has an infinite relaxation time, that is, $\tau_{\mathrm{D}} \to \infty$, also $\tau_{ii} \to \infty$. In such a case, $\mathbb{Q}_{\mathrm{cf,D}}(t)$, $\tilde{\mathbb{Q}}_{\mathrm{cf,0/d}}$, and $\mathbb{R} \to \mathbf{0}$ and, hence, $\Delta\mathbf{q}_{\mathrm{cf}}(t) = \mathbf{0}$. Therefore, as expected, only the dynamic PCM response matrix is involved in the evolution of the polarization charges, that is, $\mathbf{q}_{\mathrm{cf}}(t) = \mathbb{Q}_{\mathrm{cf,d}}\mathbf{v}_{\mathrm{M}}(t)$. Instead if the external medium has a null relaxation time, that is, $\tau_{\mathrm{D}} \to 0$, also $\tau_{ii} \to 0$. In this case,

$$\Delta\mathbf{q}_{\mathrm{cf}}(t) \equiv \mathbb{Q}_{\mathrm{cf,D}}(t)\mathbf{v}_{\mathrm{M,0}}$$

$$\to (\mathbb{Q}_{\mathrm{cf,0}} - \mathbb{Q}_{\mathrm{cf,d}})\delta(t)\mathbf{v}_{\mathrm{M,0}}$$

$$\equiv (\mathbb{Q}_{\mathrm{cf,0}} - \mathbb{Q}_{\mathrm{cf,d}})\mathbf{v}_{\mathrm{M}}(t)$$

Thus, also as expected, the entire static PCM response matrix is involved in the evolution of the polarization charges, that is, $\mathbf{q}_{\mathrm{cf}}(t) = \mathbb{Q}_{\mathrm{cf,0}}\mathbf{v}_{\mathrm{M}}(t)$.

### A.5. PCM Implementation in Octopus

Octopus is a code specialized in DFT and RT-TDDFT calculations that uses pseudopotentials to describe atomic core electrons and a real-space grid instead of basis functions.[32]

The previous PCM implementation in Octopus included only static and dynamic reaction-field polarization effects of the solvent in equilibrium with the molecular density.[37] Cavity field effects and nonequilibrium solvent polarization, either coming from the reaction or cavity field, were incorporated recently in Octopus (version 8.0) by the authors of this contribution. The propagation of the polarization charges due to solute electrons follows an EOM akin to eq 11 (see ref 13), whereas polarization charges due to solute nuclei follow a PCM equation similar to that in eq 4 (see, e.g., ref 37). Static and dynamic cavity field effects are accounted for by using eqs 4 and 11, respectively.

The EOM are integrated numerically using first-order Euler algorithm. In particular, the propagation of cavity-field polarization charges from time $t$ to $t + \Delta t$ following eq 11 is obtained from

$$\mathbf{q}_{\mathrm{cf}}(t + \Delta t) = \mathbb{Q}_{\mathrm{cf,d}}(\mathbf{v}_{\mathrm{M}}(t + \Delta t) - \mathbf{v}_{\mathrm{M}}(t)) + \Delta t\tilde{\mathbb{Q}}_{\mathrm{cf,0}}\mathbf{v}_{\mathrm{M}}(t)$$

$$- \Delta t\mathbb{R}\mathbf{q}_{\mathrm{cf}}(t) + \mathbf{q}_{\mathrm{cf}}(t) \tag{72}$$

To calculate the optical absorption spectrum with cavity field effects, our implementation in Octopus (to be released in version 9.0) relies on eq 52. $\bar{\mathbf{d}}(\omega)$ in eq 52 are calculated from eq 36a using a fast Fourier transform algorithm. $\bar{\mathbf{d}}(t)$ is computed during the real-time simulation from eq 53, the real-space approximation of eq 38b and the BEM approximation of eq 57.

Within the PCM implementation in Octopus,[37] the molecular cavity tessellation is computed using GEPOL algorithm from the union of spheres centered at each atom position except for Hydrogen.[2] The default radii of these spheres are the van der Waals radii of the corresponding atoms scaled by 1.2 (e.g., for carbon and nitrogen are 2.4 and 1.9 Å, respectively).[38] Tesserae closer than 0.1 Å are merged into a single one. To avoid singularities in the Coulomb potential resulting when grid points are close to the representative point of the cavity tessellation, we use a continuous and smooth representation of the polarization charges in each tessera (see discussion in ref 37).

### ■ AUTHOR INFORMATION

**Corresponding Authors**
*E-mail: gabrieljose.gilperez@unipd.it
*E-mail: stefano.corni@unipd.it

**ORCID** 
Gabriel Gil: 0000-0002-7728-9522
Carlo Andrea Rozzi: 0000-0002-6429-4835
Stefano Corni: 0000-0001-6707-108X

**Notes**
The authors declare no competing financial interest.

### ■ ACKNOWLEDGMENTS

We acknowledge financial support from European Research Council (ERC) under the European Union's Horizon 2020 research and innovation programme, through the project TAME-Plasmons (Grant agreement No. 681285), FP7-NMP-2011-SMALL-5 "CRONOS" (Grant No. 280879-2), and FP7-MC-IIF "MODENADYNA" (Grant No. 623413). We are grateful to A. Guandalini, for his recurrent technical assistance with Octopus installation/compilation, and J. Fregoni, for providing a preoptimized geometry of *trans*-azobenzene in vacuo using a semiempirical FOMO-SCF method.[36] We thank the members of the 'Nanostructures and (bio)molecules modeling' group from the Department of Chemistry, University of Padova, and the CNR Institute for Nanosciences, Modena, for very helpful discussions and comments.

### ■ REFERENCES

(1) Onsager, L. Electric moments of molecules in liquids. *J. Am. Chem. Soc.* **1936**, *58*, 1486–1493.

(2) Tomasi, J.; Mennucci, B.; Cammi, R. Quantum mechanical continuum solvation models. *Chem. Rev.* **2005**, *105*, 2999−3094.

(3) Cammi, R.; Mennucci, B.; Tomasi, J. On the calculation of local field factors for microscopic static hyperpolarizabilities of molecules in solution with the aid of quantum-mechanical methods. *J. Phys. Chem. A* **1998**, *102*, 870−875.

(4) Cammi, R.; Cappelli, C.; Corni, S.; Tomasi, J. On the calculation of infrared intensities in solution within the polarizable continuum model. *J. Phys. Chem. A* **2000**, *104*, 9874−9879.

(5) Tomasi, J.; Cammi, R.; Mennucci, B.; Cappelli, C.; Corni, S. Molecular properties in solution described with a continuum solvation model. *Phys. Chem. Chem. Phys.* **2002**, *4*, 5697−5712.

(6) Cammi, R.; Corni, S.; Mennucci, B.; Tomasi, J. Electronic excitation energies of molecules in solution: state specific and linear response methods for nonequilibrium continuum solvation models. *J. Chem. Phys.* **2005**, *122*, 104513.

(7) Caricato, M.; Mennucci, B.; Tomasi, J.; Ingrosso, F.; Cammi, R.; Corni, S.; Scalmani, G. Formation and relaxation of excited states in solution: A new time dependent polarizable continuum model based on time dependent density functional theory. *J. Chem. Phys.* **2006**, *124*, 124520.

(8) Cammi, R.; Tomasi, J. Time-dependent variational principle for nonlinear Hamiltonians and its application to molecules in the liquid phase. *Int. J. Quantum Chem.* **1996**, *60*, 297−306.

(9) Cammi, R.; Cossi, M.; Mennucci, B.; Tomasi, J. Analytical Hartree Fock calculation of the dynamical polarizabilities $\alpha$, $\beta$, and $\gamma$ of molecules in solution. *J. Chem. Phys.* **1996**, *105*, 10556.

(10) Liang, W.; Chapman, C. T.; Ding, F.; Li, X. Modeling ultrafast solvated electronic dynamics using time-dependent density functional theory and polarizable continuum model. *J. Phys. Chem. A* **2012**, *116*, 1884−90.

(11) Nguyen, P. D.; Ding, F.; Fischer, S. a.; Liang, W.; Li, X. Solvated First-Principles Excited-State Charge-Transfer Dynamics with Time-Dependent Polarizable Continuum Model and Solvent Dielectric Relaxation. *J. Phys. Chem. Lett.* **2012**, *3*, 2898−2904.

(12) Ding, F.; Lingerfelt, D. B.; Mennucci, B.; Li, X. Time-dependent non-equilibrium dielectric response in QM/continuum approaches. *J. Chem. Phys.* **2015**, *142*, 034120.

(13) Corni, S.; Pipolo, S.; Cammi, R. *J. Phys. Chem. A* **2015**, *119*, 5405−5416.

(14) Pipolo, S.; Corni, S.; Cammi, R. The Cavity Electromagnetic Field within the Polarizable Continuum Model of Solvation. *J. Chem. Phys.* **2014**, *140*, 164114.

(15) Jackson, J. D.; Fox, R. F. Classical electrodynamics. *Am. J. Phys.* **1999**, *67*, 841.

(16) Cancés, E.; Mennucci, B. New applications of integral equation methods for solvation continuum models: ionic solutions and liquid crystals. *J. Math. Chem.* **1998**, *23*, 309−326.

(17) Cammi, R. *Molecular Response Functions for the Polarizable Continuum Model*; Spinger: Berlin, 2013.

(18) Pipolo, S.; Corni, S. Real-Time Description of the Electronic Dynamics for a Molecule Close to a Plasmonic Nanoparticle. *J. Phys. Chem. C* **2016**, *120*, 28774−28781.

(19) Jonsher, A. K. *Dielectric Relaxation in Solids*; Chelsea Dielectrics Press, 1983.

(20) Kohanoff, J. *Electronic Structure Calculations for Solids and Molecules: Theory and Computational Methods*; Cambridge University Press, 2006.

(21) Ullrich, C. A. *Time-Dependent Density-Functional Theory: Concepts and Applications*; Oxford University Press, 2012.

(22) Kohn, W.; Sham, L. Self-Consistent Equations Including Exchange and Correlation Effects. *Phys. Rev.* **1965**, *140*, A1133.

(23) Hohenberg, P.; Kohn, W. Inhomogeneous electron gas. *Phys. Rev.* **1964**, *136*, B864.

(24) Runge, E.; Gross, E. K. Density-functional theory for time-dependent systems. *Phys. Rev. Lett.* **1984**, *52*, 997.

(25) van Leeuwen, R. Mapping from Densities to Potentials in Time-Dependent Density-Functional Theory. *Phys. Rev. Lett.* **1999**, *82*, 3863.

(26) Castro, A.; Marques, M. A. L.; Rubio, A. Propagators for the time-dependent Kohn-Sham equations. *J. Chem. Phys.* **2004**, *121*, 3425−3433.

(27) Pipolo, S.; Corni, S.; Cammi, R. The cavity electromagnetic field within the polarizable continuum model of solvation: An application to the real-time time dependent density functional theory. *Comput. Theor. Chem.* **2014**, *1040−1041*, 112−119.

(28) Novotny, L.; Hecht, B. *Principles of Nano-Optics*; Cambridge University Press, 2006.

(29) Giuliani, G.; Vignale, G. *Quantum Theory of Electron Liquid*; Cambridge University Press, 2005.

(30) Uosaki, Y.; Kitaura, S.; Moriyoshi, T. Static Relative Permittivities of Water+Ethane-1,2-diol and Water+Propane-1,2,3-triol under Pressures up to 300 MPa at 298.15 K. *J. Chem. Eng. Data* **2006**, *51*, 423−429.

(31) Ingrosso, F.; Mennucci, B.; Tomasi, J. Quantum mechanical calculations coupled with a dynamical continuum model for the description of dielectric relaxation: Time dependent Stokes shift of coumarin C153 in polar solvents. *J. Mol. Liq.* **2003**, *108*, 21−46.

(32) Andrade, X.; et al. Real-space grids and the Octopus code as tools for the development of new simulation approaches for electronic systems. *Phys. Chem. Chem. Phys.* **2015**, *17*, 31371−31396.

(33) Schelter, I.; Kümmel, S. Accurate Evaluation of Real-Time Density Functional Theory Providing Access to Challenging Electron Dynamics. *J. Chem. Theory Comput.* **2018**, *14*, 1910−1927.

(34) Osborn, J. Demagnetizing factors of the general ellipsoid. *Phys. Rev.* **1945**, *67*, 351.

(35) Böttcher, C. J. F.; van Belle, O. C.; Bordewijk, P.; Rip, A. *Theory of Electric Polarization*; Elsevier: Amsterdam, 1973; Vol. 1.

(36) Fregoni, J.; Granucci, G.; Coccia, E.; Persico, M.; Corni, S. Manipulating azobenzene photoisomerization through strong light-molecule coupling. *Nat. Commun.* **2018**, *9*, 4688.

(37) Delgado, A.; Corni, S.; Pittalis, S.; Rozzi, C. A. Modeling solvation effects in real-space and real-time within density functional approaches. *J. Chem. Phys.* **2015**, *143*, 144111.

(38) Cancés, E.; Mennucci, B.; Tomasi, J. A new integral equation formalism for the polarizable continuum model: Theoretical background and applications to isotropic and anisotropic dielectrics. *J. Chem. Phys.* **1997**, *107*, 3032.